\documentstyle[prd,aps]{revtex}

\begin{document}
\tightenlines
\parskip 0.3cm

\begin{centering}
{\large \bf On Neutrino Absorption Tomography of the Earth}\\
\vspace{.9cm}
{ Matias M. Reynoso and Oscar A. Sampayo}\\
\vspace{.05in}
{\it  Departamento de F\'{\i}sica,
Universidad Nacional de Mar del Plata \\
Funes 3350, (7600) Mar del Plata, Argentina} \\ \vspace{.4cm}
 \vspace{.05in}
{ \bf Abstract}
\\
\bigskip
\end{centering}
{\small We study the passage of UHE neutrinos through the Earth in
order to perform an absorption tomography of its inner structure.
The aim of this work is to study the extraction methods of the
Earth's density, in this conditions, we do not need to implement a
realistic Monte Carlo simulation, as we are only interested in
comparing the goodness of a standard method \cite{ralston} with
the one we propose. The Earth's density is reconstructed using the
2-d Radon transform and we compare the density obtained
considering neutral current regeneration through the complete
transport equation, with the one obtained making use of the
effective cross section approximation (standard method). We see
that the effective cross section leads in general to inaccurate
results, especially for flat initial neutrino fluxes, while the
full transport equation method works regardless of the initial
flux. Finally, an error propagation analysis made for different
uncertainties in the surviving neutrino flux shows that the
recovered density presents a percentage uncertainty less than two
times the uncertainty in the flux. }

\pacs{PACS: 96.40.Tv, 91.35.Pn}

\vspace{0.2in} 


\section{\bf Introduction}
Neutrino absorption tomography has been considered as an
alternative way to obtain information about the interior of the
Earth that would be independent of any geophysical model (see
\cite{ralston} and references therein). In view of this, it is
useful to understand the role of neutral current regeneration of
neutrinos in their passage though the Earth. To that end, in the
present work we perform the absorption tomography following two
different approaches to take account of the neutral current
regeneration effect in the density extraction method. The first
approach, which we call the standard method or the effective cross
section approach  \cite{ralston2}, is largely used in the
literature consists on implementing an approximate solution to the
transport equation by defining an effective cross section. The
other method makes use of the complete integro-differential
transport equation itself and not of its solution, and this is the
approach we introduce in this work.



In the next section we start by describing both the standard and
the transport equation approaches, and apply them for some initial
neutrino fluxes. We use theorized neutrino fluxes in the range
$10^4$GeV to $10^8$GeV coming from cosmic sources such as Active
Galactic Nuclei (AGN) and Gamma Ray Bursts (GRB) as well as
neutrinos generated by cosmic ray interactions in the atmosphere
(ATM). All these fluxes are assumed to be isotropic, and the Earth
density is considered spherically symmetric so that the angular
data of the surviving flux can be used to recover the density be
means of the 2-d Radon transform which is a standard technique,
but as we have not found it explained in any article, we decided
to include it in the appendix. In order to have a surviving
neutrino flux after traversing the Earth, we consider the
numerical solution to the complete transport equation for neutrino
propagation. This topic is presented in section 2 and the
reconstruction procedures are discussed in section 3, where we
compare the results of our approach with the ones of the
$\sigma_{eff}$-approach. For the sake of comparing the two density
extraction methods, it suffices to consider the numerical solution
for the surviving flux, since a realistic Monte Carlo study would
be consistent with the numerical results for the surviving flux.
Finally, in section 4 the error propagation is performed when
recovering the Earth's density following the transport equation
approach, that is making no approximation related with the NC
regeneration.

\section{\bf Neutrino Propagation Through the Earth.
Transport Equation and Effective Cross Section}

As neutrinos travel through the Earth they may suffer charged
current (CC) and neutral current (NC) interactions with the
nucleons in their path. Neutrino oscillation within the Earth can
be neglected for energies higher than $1000$GeV \cite{stasto}, and
since the CC and NC cross sections increase with the neutrino
energy, we choose our energy range from $10^4$GeV to $10^8$GeV
where the Earth is neither totally transparent nor opaque to
neutrinos.

The change in a neutrino flux $\phi_\nu(E)=\frac{dN}{dE dt}$ as it
traverses the Earth can be divided into two effects: absorption
and regeneration \cite{nicolaidis}. Absorption is a decrease in
the neutrino flux due to CC or NC interactions
When neutrinos pass through an amount of matter $d\tau= n(r) dz$
in a distance $dz$, where $n(r)$ is the Earth's number density,
the change in the flux $\phi_\nu(E,\tau)$ due only to absorption
is proportional to $\phi_\nu(E,\tau)$ by the total cross section
$\sigma_{tot}(E)=\sigma_{cc}(E)+\sigma_{nc}(E)$ which represents a
probability of CC or NC interaction:
\begin{eqnarray}\label{asu0}
\frac{d\phi_\nu(E,\tau)}{\hspace{-0.6cm}d\tau}=-\sigma_{tot}(E)
\phi_\nu(E,\tau)
\end{eqnarray}
Here $\tau(z)$ is the amount of material found until a depth $z$
that is,
\begin{eqnarray}\label{tau}
\tau(z)  = \int^z_0 dz' n(z')
\end{eqnarray}
where the number density is the Avogadro's constant times the
density, $n(z')=N_A \rho(z')$.

To write the complete transport equation for neutrinos of our
energy range, we have to add to (\ref{asu0}) the effect of
regeneration, which accounts for the possibility that neutrinos of
energies $E'>E$ may end up with energy $E_\nu$ due to NC
interactions with the nucleons.
Unlike absorption, regeneration represents an increase in the flux
at an energy $E$ caused by NC interactions at energies $E'>E$.
Now, this change due to regeneration is proportional to each flux
$\phi_\nu(E',\tau)$ by the probability of such interactions which
is given by $\frac{d\sigma}{dE}(E',E)$ so that we can write the
transport equation as follows
\begin{eqnarray}\label{eqtras}
\frac{d\phi_\nu(E,\tau)}{\hspace{-0.6cm}d\tau}=-\sigma_{tot}(E)
\phi_\nu(E,\tau) + \int_E^\infty dE' \
\frac{d\sigma_{nc}(E',E)}{\hspace{-0.6cm}dE} \ \phi_\nu(E',\tau).
\end{eqnarray}
to be solved with the initial condition
$\phi_\nu(E,0)=\Phi_\nu(E)$, where $\Phi_\nu(E)$ is the initial
flux.

The other approach consists on defining an effective cross section
$\sigma_{eff}(E,\Phi_\nu)$ in an attempt to incorporate both
absorption and regeneration all together. Dividing the transport
equation (\ref{eqtras}) by $\phi_\nu(E,\tau)$ we obtain
\begin{eqnarray} \label{eqdiv}
\frac{d \ln
\phi_\nu(E,\tau)}{\hspace{-0.8cm}d\tau}=-\sigma_{tot}(E)
 + \int_{E_\nu}^\infty dE' \
\frac{d\sigma_{nc}(E',E)}{\hspace{-0.6cm}dE} \
\frac{\phi_\nu(E',\tau)}{\phi_\nu(E,\tau)}.
\end{eqnarray}
In this last expression, the quotient of the fluxes
$\frac{\phi_\nu(E',\tau)}{\phi_\nu(E,\tau)}$ appears integrated on
$E'$, and these fluxes must be evaluated at a depth in the Earth
corresponding to $\tau$ nucleons traversed. The effective cross
section approach consists in replacing the above quotient by the
initial flux ratio \cite{ralston2}
\begin{eqnarray}\label{apro}
\frac{\phi_\nu(E',\tau)}{\phi_\nu(E,\tau)} \longrightarrow
\frac{\Phi_\nu(E')}{\Phi_\nu(E)},
\end{eqnarray}
and now we identify the effective cross section as
\begin{eqnarray} \label{seff}
\sigma_{eff}(E,\Phi_\nu)=-\sigma_{tot}(E)
 + \int_{E_\nu}^\infty dE' \
\frac{d\sigma_{nc}(E',E)}{\hspace{-0.6cm}dE} \
\frac{\Phi_\nu(E')}{\Phi_\nu(E)}.
\end{eqnarray}
The effective cross section results then from the approximation
(\ref{apro}) to use the initial fluxes (which are supposed as
known) rather than the fluxes $\phi_\nu(E,\tau)$ at different
depths through the Earth which can be obtained using the original
transport equation (\ref{eqtras}). When using the effective cross
section, the resulting flux for an amount of material $\tau$
traversed is
\begin{eqnarray} \label{flueff}
\phi_\nu(E,\tau)=\Phi_\nu(E)e^{-\sigma_{eff}(E) \tau}
\end{eqnarray}
We can compare this prediction for the surviving flux with the one
obtained by solving the transport equation numerically. To do
this, we use the different initial diffuse fluxes for: ATM
neutrinos derived by Volkova \cite{volkova}, neutrinos coming from
AGN according to Stecker-Salomon \cite{stecker} and to Protheroe
\cite{protheroe}, and neutrinos from GRB derived by Waxman
\cite{waxman} (see figure 1).

Besides the initial flux, we also need to assume a density for the
Earth to solve the transport equation or to use the
$\sigma_{eff}$-approach, so we take this density to be given by
the Preliminary Reference Earth Model (PREM) \cite{premm}, which
is plotted in figure 2. Neutrinos will then  follow a path through
the Earth towards the detector as illustrated in figure 3.


 Solving the transport equation by Euler's method and by
the iterative method described in \cite{R1} give the same results
which are also consistent with the ones in \cite{stasto}. For
illustration we show in figures 4 and 5 the normalized final flux
or shadowing factor
\begin{eqnarray}
S(E,\theta)=\frac{\phi_\nu(E,\tau)}{\Phi_\nu(E)}
\end{eqnarray}
as well as the shadowing factor $S_{eff}$ according to the
$\sigma_{eff}$ approach
\begin{eqnarray}\label{sheff}
S_{eff}(E,\theta)=e^{-\sigma_{eff}(E,\Phi_\nu) \tau}
\end{eqnarray}
for nadir angles $\theta=\lbrace 80^0, 40^0 ,0^0 \rbrace$ from the
downward normal to the surface where the detector is (figure 3).

We note that the two approaches agree when dealing with fast
decreasing initial spectra such as ATM, while for flatter initial
fluxes, the $\sigma_{eff}$ approach trends to overestimate the
surviving flux, (and hence the shadowing factor) as the nadir
angle decreases.
\newpage

\section{\bf Tomography of the Earth}

In the appendix we see that the Radon transform of the number
density is the amount of material traversed $\tau$ appearing in
the transport equation, and the expressions for the Radon
transform and its inverse transform are obtained exploding the
assumed spherical symmetry of the density. For reconstructing the
number density we will then use (see the appendix)
\begin{eqnarray}\label{reco}
n(r)=-\frac{1}{\pi} \int_{\arcsin \frac{r}{R}}^{\pi/2} \frac{d
\theta'}{ \sqrt{R^2 \sin^2 \theta'-r^2}}
\frac{d\tau(\theta')}{d\theta'}
\end{eqnarray}
This expression relates the number density of the Earth at a
distance $r$ from its center, with the different amounts of matter
traversed at different nadir angles through the Earth. Then,
measuring $\tau(\theta')$ at all the possible nadir angles between
$0$ y $\pi/2$ we may obtain $n(r)$.

To use equation (\ref{reco}), we must be able to infer
$\frac{d\tau}{d\theta}$ from the data of the surviving neutrino
flux. We may then do this inference under the transport equation
approach or under the $\sigma_{eff}$-approach and reconstruct a
density in each case.

Leaving for the next section a treatment of the error involved, we
now generate data for the surviving flux at the $91$ nadir angles
${90^0,89^0,...,0^0}$ as we did in the previous section for
$(80^0,40^0,0^0)$. These data are generated using the complete
transport equation, assuming the density given by the PREM (figure
2), and now we consider two different initial spectra for the sum
of ATM, AGN, and GRB neutrinos (figure 6).

Once we have the surviving neutrino flux, we have to obtain
$\frac{d\tau(\theta)}{d\theta}$ in order to do the tomography of
the Earth. The usual approach is to use the effective cross
section to relate the surviving flux with $\tau(\theta)$ through
(\ref{flueff}), but as we pointed out in the previous section,
this approach is only satisfactory for very fast decreasing
spectra. In this conditions, we expect that the reconstructed
density may not be accurate in general.

As an alternative method, we propose to use the complete
integro-differential transport equation (not its solution) without
approximation to directly obtain $\frac{d\tau(\theta)}{d\theta}$
from the surviving flux $\phi_\nu(E,\theta)$, which can be done
noting that
\begin{eqnarray}\label{chain}
\frac{d\phi_\nu(E,\theta)}{\hspace{-0.6cm}d\theta}=\frac{d\phi_\nu(E,\tau(\theta))}{\hspace{-0.8cm}d\tau}
\cdot \frac{d\tau}{d\theta}.
\end{eqnarray}
As the  angular variation of the number of nucleons $\tau$ is
independent of the energy of the neutrino flux, if we integrate
over the our neutrino energy range $\Delta
E=(10^4$GeV$,10^7$GeV$)$ at both sides of (\ref{chain}) and use
the transport equation (\ref{eqtras}), we find that
\begin{equation}\label{dtauTE}
\frac{d\tau(\theta)}{\hspace{-0.2cm}d\theta}=\frac{N(\theta)}{D(\theta)}
\end{equation}
where
\begin{eqnarray}
 N(\theta)\hspace{-0.2cm}&=&\hspace{-0.3cm}\int_{\Delta E} dE \
\frac{d\phi_\nu (E,\theta)}{\hspace{-0.6cm}d\theta} \nonumber
\\ D(\theta)\hspace{-0.2cm}&=& \hspace{-0.3cm}\int_{\Delta E} dE \left( \int_E^\infty dE' \
\frac{d\sigma_{nc}(E',E)}{\hspace{-0.6cm}dE} \
\phi_\nu(E',\theta)-\sigma_{tot}(E)
\phi_\nu(E,\theta)\right).\nonumber
\end{eqnarray}
Inserting $\frac{d\tau(\theta)}{\hspace{-0.2cm}d\theta}$ given by
(\ref{dtauTE}) into (\ref{reco}) yields the density reconstructed
under the transport equation approach.


The densities recovered are plotted in figure 7, together with the
original one given by the PREM. We would like to point out again
that we reconstruct the earth density using the differential
transport equation. We do not need to use the corresponding
solution. As the differential transport equation is locally valid
 we have considered it and the corresponding
measurable surviving flux on the earth surface.

We clearly observe that the recovered density
using the effective cross section approach is higher that the
density that actually caused the attenuation, while with the
transport equation approach the reconstructed density acceptably
reproduces the original one. We note that the effective cross
section leads to worse results the flatter is the initial flux, as
we saw in the previous section. This is due to the fact that NC
regeneration is more important when there are "many" neutrinos of
higher energy than a given one, that then may happen to regenerate
to that lower energy. With a fast-decreasing spectra such as ATM
alone, regeneration is not very important, and hence the
$\sigma_{eff}$-approach also works for recovering the original
density. However, in our range of energy where absorption
tomography is in principle possible,  other sources of neutrinos
also are expected to contribute (AGN, GRB, etc.) and dominate in a
total neutrino flux in that range. This justifies the use of the
complete transport equation approach rather than the standard one.

\section{Error Propagation}
In this section we present the error propagation analysis
considering only the uncertainty in the surviving neutrino flux.
We have not taken into account theoretical uncertainties in the
cross section such as the error due to extrapolating  to high
energies, or the uncertainty accounting for possible new physics.

The method we suggest for performing a neutrino absorption
tomography clearly depends on the surviving neutrino flux at each
nadir angle. IceCube \cite{icecu} in the southern ice seems to be
the most promising neutrino detector, with $1km^3$ of detection
volume with photomultipliers placed regularly to catch the
Cherenkov light emitted when CC produced muons travel in the ice.
The angular resolution is expected to be of about $1^0$ (or
better), which is the angular resolution we implemented for doing
the density extraction.

As the uncertainty in a total measured neutrino flux depends among
other things on the surviving flux being measured, we do not know
what it will be for certain. Still, to test the recovery
procedure, we assume three different uncertainties in the detected
flux which are in the order of the expected ones in
\cite{icecu,incert}, and propagate each uncertainty to obtain the
corresponding one to the recovered density.
To do this, we implement the transport equation approach,
in which $\frac{d\tau(\theta)}{\hspace{-0.2cm}d\theta}$ is given
by (\ref{dtauTE}).
So for an uncertainty in the flux $\Delta \phi_\nu(E,\theta)$ we
find that the uncertainties in $N(\theta)$ and $D(\theta)$ are
given by
\begin{eqnarray}
\Delta N(\theta)&=&\frac{d}{d\theta}\int_{\Delta E} dE \Delta
\phi_\nu(E,\theta) \\ \Delta D(\theta)&=& \sqrt{(\Delta
d_1(\theta))^2+(\Delta d_2(\theta))^2} ,
\end{eqnarray}
where
\begin{eqnarray}
\Delta d_1(\theta)&=&\int_{\Delta E} dE \sigma_{tot}(E) \Delta
\phi_\nu(E,\tau)\nonumber\\ \Delta d_2(\theta)&=&\int_{\Delta E}
dE \int_E^\infty  \ dE' \
\frac{d\sigma_{nc}(E',E)}{\hspace{-0.6cm}dE} \ \Delta
\phi_\nu(E',\tau)\nonumber.
\end{eqnarray}
The absolute error in $\frac{d\tau}{d\theta}$ is then
\begin{eqnarray}
\Delta
\left[\frac{d\tau}{d\theta}(\theta)\right]=\sqrt{\left(\frac{\Delta
N(\theta)}{D(\theta)}\right)^2+\left(\frac{N(\theta)}{D(\theta)}\Delta
D(\theta)\right)^2},
\end{eqnarray}
and by means of (\ref{reco}), we obtain the uncertainty in the
number density recovered:
\begin{eqnarray}\label{unreco}
\Delta n(r)=-\frac{1}{\pi} \int_{\arcsin \frac{r}{R}}^{\pi/2}
\frac{d \theta'}{ \sqrt{R^2 \sin^2  \theta' -r^2}}
\left(\Delta\left[\frac{d\tau(\theta')}{d\theta'}\right]\right)
\end{eqnarray}
that yields the corresponding uncertainty in the density through
$\Delta \rho(r)=\Delta n(r)/N_A$, which we illustrate by error
bars in the figures 8 and 9 for the three uncertainties
 of $15\%$, $10\%$, and $5\%$ in the surviving neutrino
flux. We show the results with the error bars considering only the
initial flux produced by AGN (Stecker-Salomon), GRB and ATM
neutrinos added together, as they are substantially the same the
other initial spectra considered before (AGN (Protheroe) + GRB +
ATM) under the transport equation approach.

We observe that the error propagated to the recovered density is
about $1.6$ - $1.7$ times the error in the flux detected.

\section{Final Remarks} 
We have presented in this work a simple method to extract the
Earth's density based on the complete transport equation, without
using any approximation as the use of the effective cross section
would imply. We have compared the results obtained by our method
with the ones obtained by the $\sigma_{eff}$-method. We can
conclude that our method allows an accurate reconstruction for
each initial flux we have considered, while the standard approach
trends to overestimate the density accounting for its inaccuracy
in dealing with the NC regeneration.

To estimate the uncertainties in the reconstructed density, we
have propagated the relative uncertainty in the surviving neutrino
flux and found that it is increased about 60\% - 70\% by the
procedure.


\appendix
    \section{Radon Transform and the Earth's density}
The 2-d Radon transform \cite{radoncross} of a continuous function
$n(x,y)$ can be defined as
\begin{equation} \label{radon}
R_{n}(p,\vec{\xi})=\int_{\Re^{2}} \ dx \ dy \ n(x,y) \
\delta(p-\vec{x}\cdot\vec{\xi})
\end{equation}

$R_{n}(p,\vec{\xi})$ (see figure 10) is thus $n(x,y)$ integrated
along the straight at a distance $p$ from the origin and normal to
$\vec{\xi}=(\cos \theta ,\sin \theta)$ which is an unitary vector
defined by $ \vec{\xi}=(\cos \theta ,\sin \theta)$,
where the angle $\theta$ is measured from the x-axis (Fig.3).

 If $R_{n}(p,\vec{\xi})$ is continuous, then $n(x,y)$ can be
recovered using the inverse Radon transform. The latter can be
obtained relating the 2-d Radon transform with the 2-d Fourier
tranform $F_{n}(\lambda \vec{\xi})$
\begin{equation}\label{fouri}
F_{n}(\lambda \vec{\xi})= \int_{\Re^{2}}dx \ dy \ e^{i\lambda
\vec{x}\cdot\vec{\xi}}n(\vec{x}),
\end{equation}
where $\vec{x}=(x,y)$.

 Since $\int_{-\infty}^{\infty}dp \ \delta(p-
\vec{x}\cdot\vec{\xi})=1$,
\begin{equation}\label{fourin}
 F_{n}(\lambda \vec{\xi})= \int_{-\infty}^{\infty} \ dp
\delta(p-\vec{x}\cdot\vec{\xi})\int_{\Re^{2}}dx \ dy \
e^{i\lambda\vec{x}\cdot\vec{\xi}}n(\vec{x})\nonumber
\end{equation}
and as the Dirac's delta  $\delta(p-\vec{x}\cdot\vec{\xi})$ allows
us to write $p$ instead of $\vec{x}\cdot\vec{\xi}$, we obtain
\begin{eqnarray}
  F_{n}(\lambda\vec{\xi})=\int_{-\infty}^{\infty} dp \
  e^{i\lambda \ p}\int_{\Re^{2}}dx
\ dy \ \delta(p-\vec{x}\cdot\vec{\xi}) \ n(\vec{x}) \label{A6}.
\end{eqnarray}
That is, the 2-d Fourier transform of $n(\vec{x})$ is equal to the
2-d Fourier transform of the 2-d Radon transform of $n(\vec{x})$,
which is known as the Fourier Slice Theorem:
\begin{equation}
  F_{n}(\lambda\vec{\xi})=\int_{-\infty}^{\infty} dp \
e^{i\lambda \ p}\ R_{n}(p,\vec{\xi}) \label{A7}.
\end{equation}
Then, by means of the inverse Fourier transform we recover
$n(\vec{x})$,
\begin{eqnarray}
n(\vec{x})=\frac{1}{(2\pi)^{2}}\int d^{2}(\lambda \vec{\xi}) \
e^{-i\lambda \vec{x}\cdot\vec{\xi}} F_{n}(\lambda \vec{\xi})
\label{A8}
\end{eqnarray}
Noting that the unitary vector $\vec{\xi}=(\cos \ \phi,\sin \
\phi)$,
we use the polar coordinates $(\lambda,\phi)$ and (\ref{A8})
becomes
\begin{eqnarray}
n(\vec{x})=\frac{1}{(2\pi)^{2}}\int_{0}^{\infty}d\lambda \ \lambda
\int_{0}^{2\pi}d\phi \  e^{-i\lambda \vec{x}\cdot\vec{\xi}}
F_{n}(\lambda\vec{\xi}) \label{A9}.
\end{eqnarray}
Expressing $F_{n}(\lambda\vec{\xi})$ according to (\ref{A7}),
defining $p':=p-\vec{x}\cdot\vec{\xi}$, and
$\overline{R}_{n}(p',\vec{\xi})$ as
\begin{equation}
\overline{R}_{n}(p',\vec{\xi}):=\frac{1}{2\pi}
\int_{0}^{2\pi}d\phi \ R_{n}(p'+\vec{x}\cdot\vec{\xi},\vec{\xi})
\label{A12}
\end{equation}
we can write $n(\vec{x})$ as
\begin{equation}
n(\vec{x})=\frac{1}{2\pi}\int_{0}^{\infty}\lambda \ d\lambda
\int_{-\infty}^{\infty}dp \ e^{i\lambda
p}\overline{R}_{n}(p,\vec{\xi}) \label{A13}.
\end{equation}

Expressing the above equation in terms of the function
$sign(\lambda)$, performing integration by parts in $p$, and using
the Fourier transform of $sign (\lambda)$, we obtain
\begin{equation}
n(\vec{x})=-\frac{1}{2\pi}\int_{-\infty}^{\infty}dp
\frac{\overline{R}'_{n}(p,\vec{\xi})}{p} \label{A19}
\end{equation}
which expresses $n(\vec{x})$ in terms the $p$ derivative of its
Radon transform in a simplified fashion.



In the following, we use the above results to relate the Earth's
density with its Radon transform in a simple expression.

 We assume an spherically symmetric number density for
the Earth which we denote by $n(r)$ and begin by noting that the
amount of nucleons $\tau(z)$ found along a path of depth $z=2R
\cos \theta$ in the Earth is nothing but the Radon transform of
the number density $n(r)$ of the Earth:
\begin{eqnarray}\label{taura}
\tau(\theta) = \int^z_0 dz' n(z') = \int_{-\infty}^\infty dx
\int_{-\infty}^\infty dy \ n(\vec{x}) \ \delta(p- \vec{\xi}\cdot
\vec{x})
\end{eqnarray}
where now $p=R \sin \ \theta$ and $R$ is the Earth's radius.
In the polar coordinates $(r,\phi)$ with the origin in the Earth's
center, $\vec{\xi}\cdot \vec{x} = r \cos (\phi-\theta)$, then we
may write
\begin{eqnarray}\label{nueve}
\tau(\theta) = \int_0^{2\pi} d\phi \int_0^\infty dr \ r \ n(r) \
\delta(p- r \cos (\phi-\theta)).
\end{eqnarray}
We can now integrate in $\phi$ making use of the Dirac's delta
property $\delta(f(\phi))=\sum_i \frac{\delta(\phi - \phi_i)}{ |
\frac{df}{d\phi}|_{\phi_i}}$ where $\phi_i$ are the zeros of
$f(\phi)$, and as this Dirac's delta in (\ref{nueve}) implies that
\begin{eqnarray}
\cos(\phi_i-\theta)=\frac{p}{r}  \hspace{1cm} \mbox{then }
\hspace{0.5cm } |p|<r \nonumber,
\end{eqnarray}
and expression (\ref{nueve}) becomes
\begin{eqnarray} \label{once}
\tau(\theta) = \int_{|p|}^\infty \frac{2r \ n(r) \
dr}{\sqrt{r^2-p^2}}.
\end{eqnarray}
Then as $n(r)$ vanishes for $r>R$,
\begin{eqnarray} \label{Rn}
\tau(\theta) = R_n(p)= \int_{R^2 \sin^2 \theta}^{R^2} \frac{n(r) \
d(r^2)}{\sqrt{r^2- R^2 \sin^2\theta}}.
\end{eqnarray}
We can recover $n(r)$ by means of the expression for the inverse
Radon transform (\ref{A19}):
\begin{eqnarray}
n(r)=-\frac{1}{(2\pi)^2} \int_{-\infty}^\infty \frac{dp}{p} \
\frac{d}{dp} \int_0^{2\pi} d\phi \ R_n(p+\vec{\xi} \cdot \vec{x})
\end{eqnarray}
defining $p':=p+\vec{\xi} \cdot \vec{x}$,
we can write
\begin{eqnarray}
n(r)=-\frac{1}{(2\pi)^2} \int_{-\infty}^\infty \int_0^{2\pi}
\frac{dp' \ d\phi}{p'-r \ \cos(\phi-\theta)} \
\frac{dR_n(p')}{dp'}.
\end{eqnarray}
Now, the integral in $\phi$ is defined for $p'>r$ and $p'<-r$,
then by the Cauchy theorem it turns out that
\begin{eqnarray}
\int_0^{2\pi} \frac{d\phi}{p'-r \cos(\phi-\theta)} = \left
\lbrace\begin{array}{c l} \frac{2\pi}{p'^2 - r^2}  \ \hspace{1cm}
\mbox{si }p'>r \vspace{0.3cm}\\ -\frac{2\pi}{p'^2 - r^2}  \
\hspace{1cm} \mbox{si }p'<-r
\end{array}\right. \nonumber
\end{eqnarray}
Then, as it must be $|p'|<R$ so that $R_n(p')$ does not vanish,
defining $\theta'$ such that $p'=R \sin \ \theta'$, and recalling
that $R_n(p')=\tau(\theta')$, we obtain that
\begin{eqnarray}\label{rec}
n(r)=-\frac{1}{\pi} \int_{\arcsin \frac{r}{R}}^{\pi/2} \frac{d
\theta'}{ \sqrt{R^2 \sin^2  \theta'-r^2}}
\frac{d\tau(\theta')}{d\theta'},
\end{eqnarray}
which allows us to express the Earth's density in terms of the
slope in the amount of nucleons $\tau$.


{\bf Acknowledgements}

We thank CONICET (Argentina) and Universidad Nacional de Mar del
Plata (Argentina) for their financial supports.



\noindent{\large \bf Figure Captions}\\

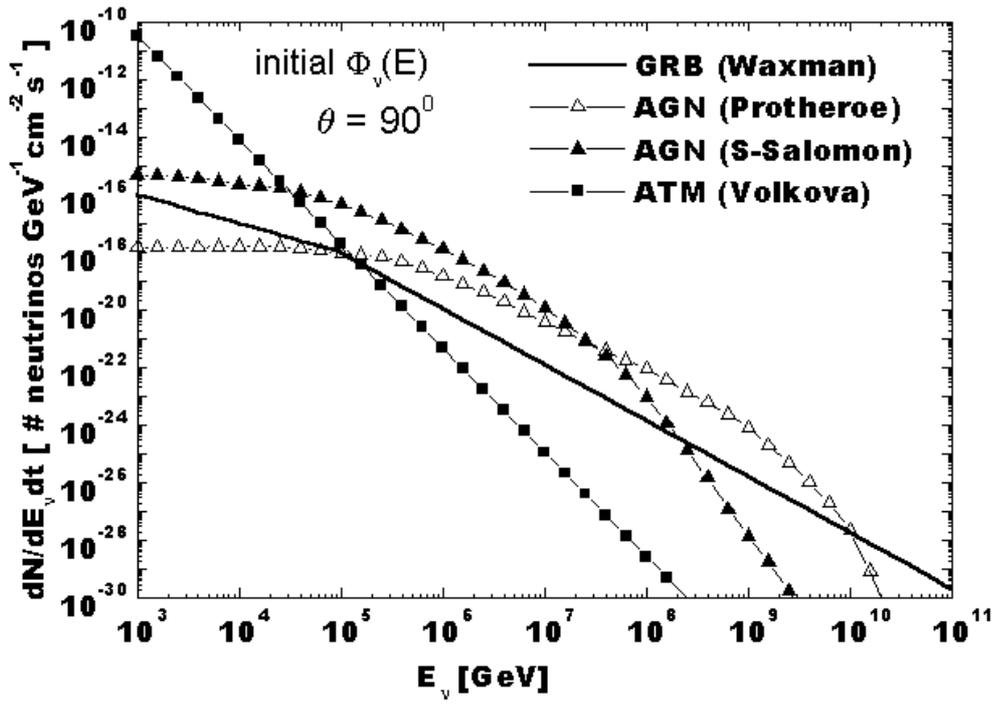
\noindent{\bf Figure 1:} Initial muon neutrino fluxes plotted
separately: AGN(Protheroe) \cite{protheroe}, AGN(Stecker-Salomon)
\cite{stecker}, GRB \cite{waxman} and ATM \cite{volkova}.

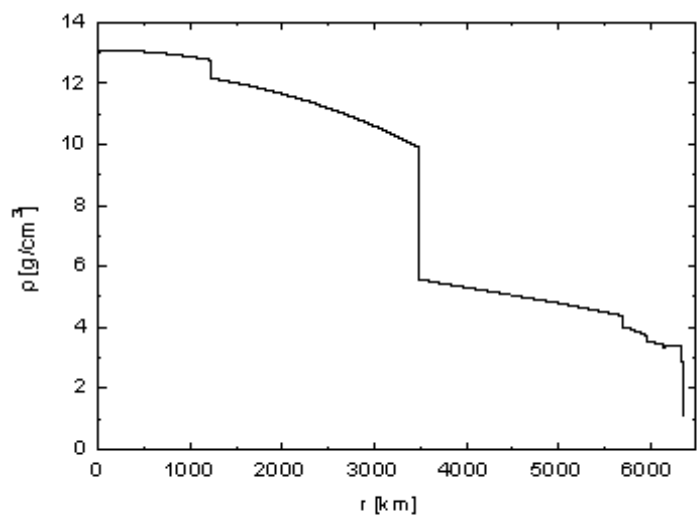
\noindent{\bf Figure 2:} Earth density as predicted by the
'Preliminary Reference Earth
        Model' \cite{premm}.

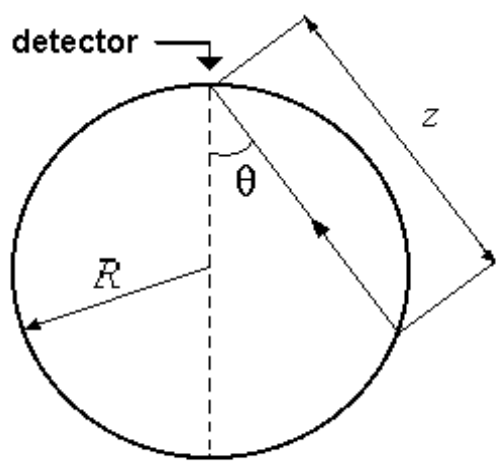
\noindent{\bf Figure 3:} Neutrino path towards the detector.

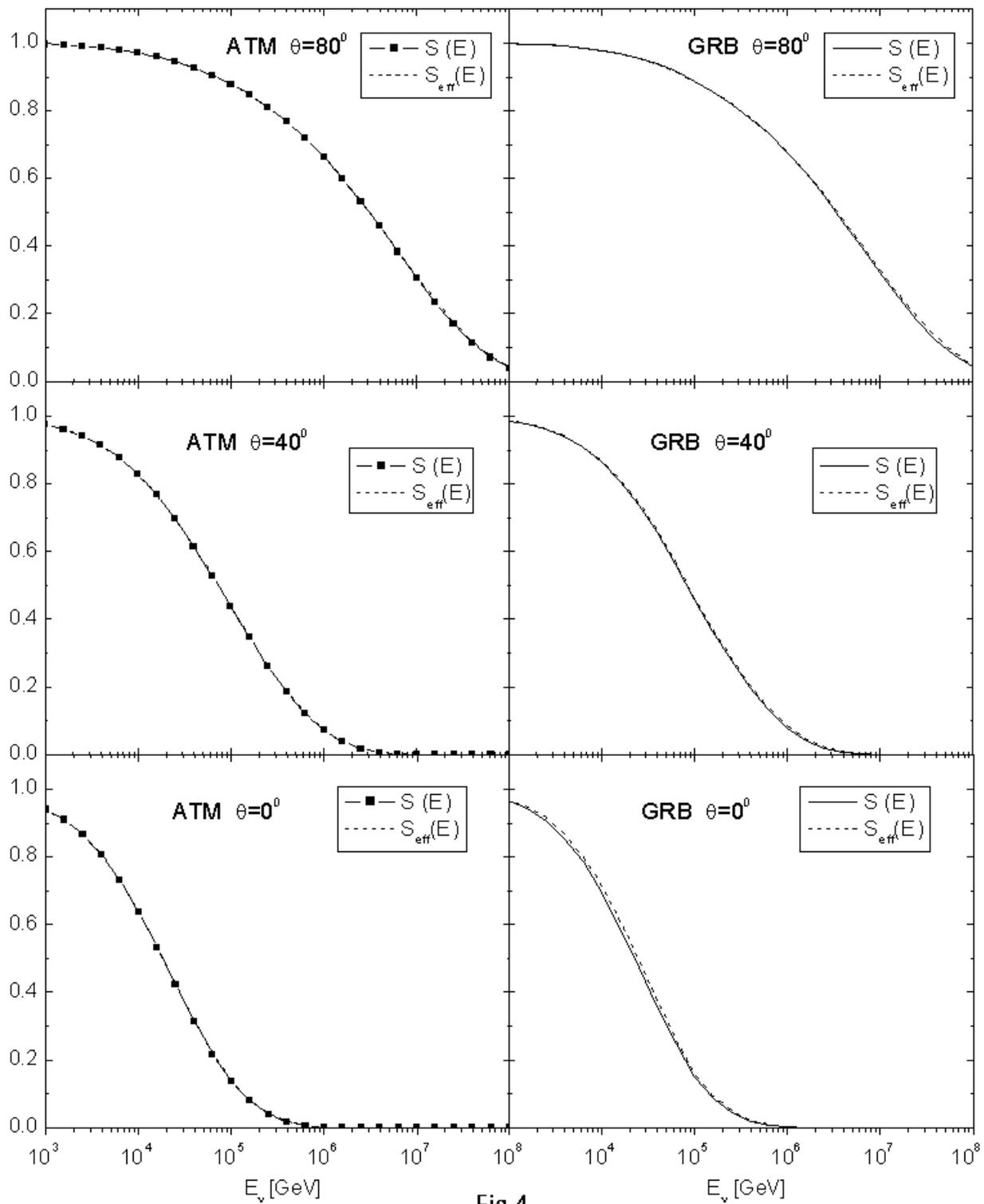
\noindent{\bf Figure 4:} Shadowing factors for ATM and GRB fluxes
for nadir angles $80^0$, $40^0$, and $0^0$.

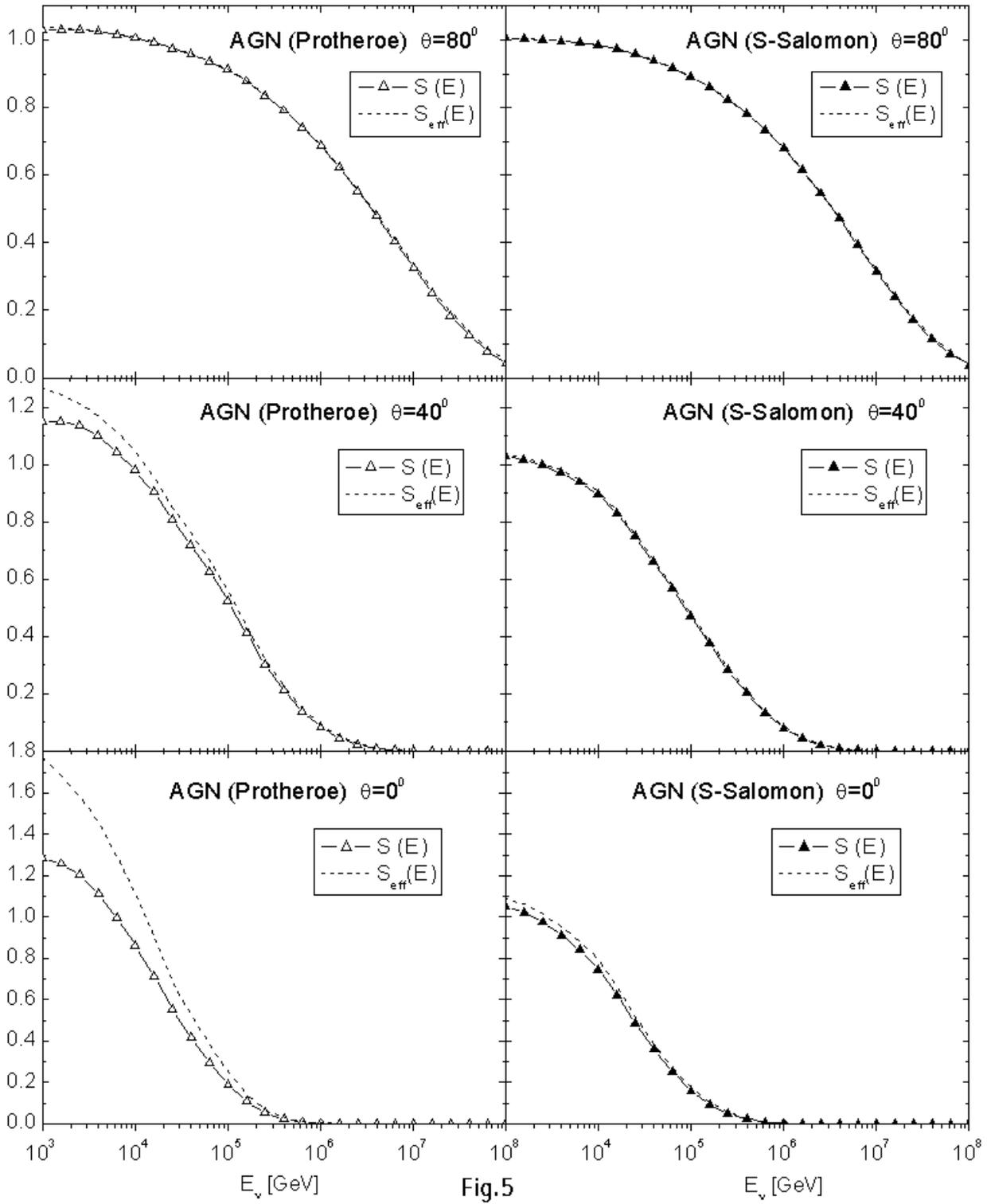
\noindent{\bf Figure 5:} Shadowing factors for AGN (S-Salomon) and
AGN (Protheroe) fluxes for nadir angles $80^0$, $40^0$, and $0^0$.

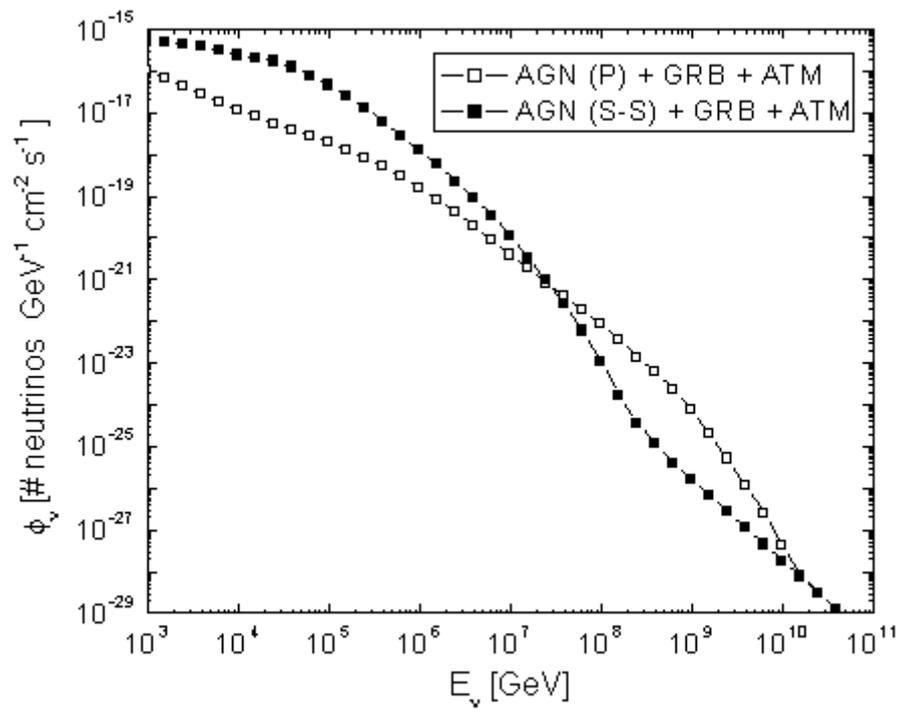
\noindent{\bf Figure 6:}  Total initial neutrino fluxes from AGN +
GRB + ATM. AGN(P) stands for the flux by Protheroe, and AGN(S-S)
       is the neutrino flux predicted by Stecker and
       Salomon.

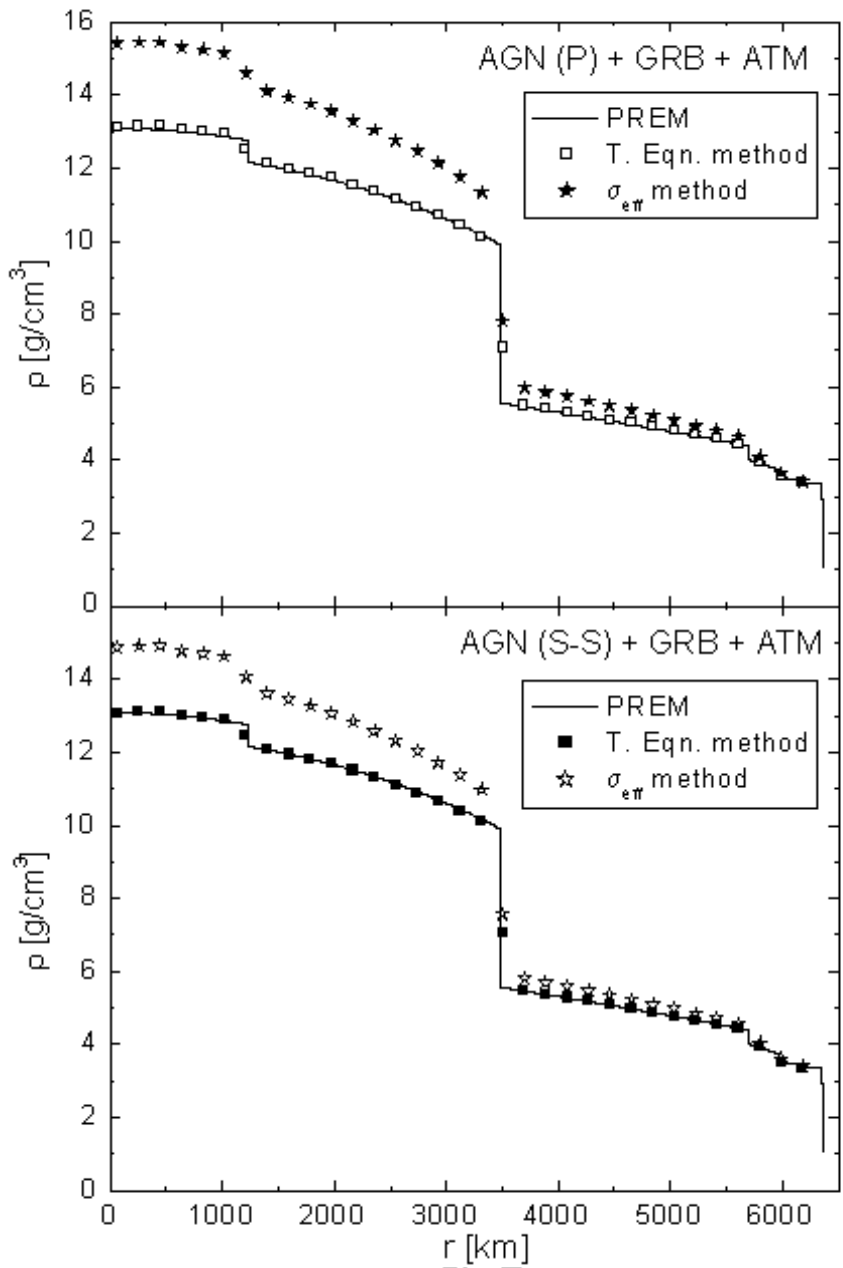
\noindent{\bf Figure 7:} Recovered densities compared with the
PREM density (solid line). Boxes and stars represent the densities
obtained with the transport equation and with the $\sigma_{eff}$
methods respectively.

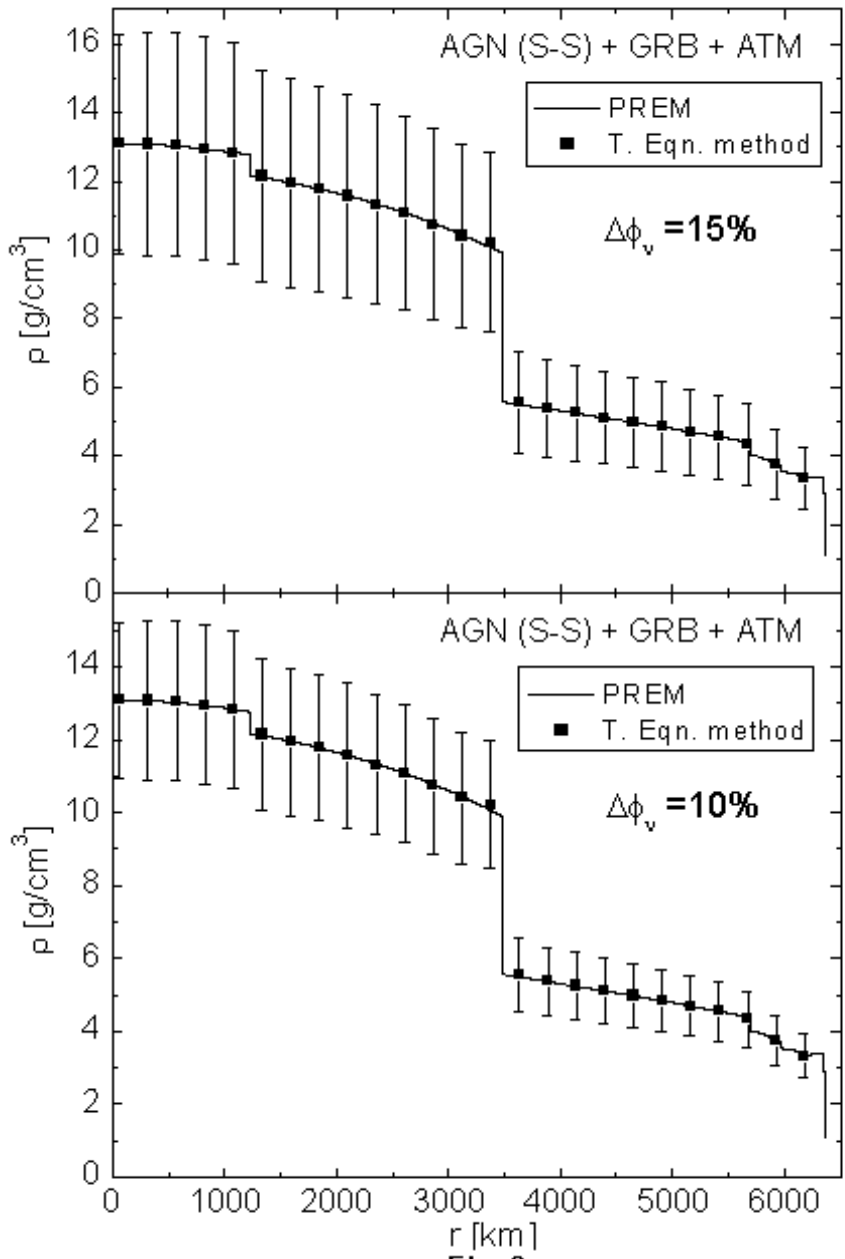
\noindent{\bf Figure 8:} Densities recovered with error bars for
uncertainties of $15\%$ and $10\%$ in the surviving flux.

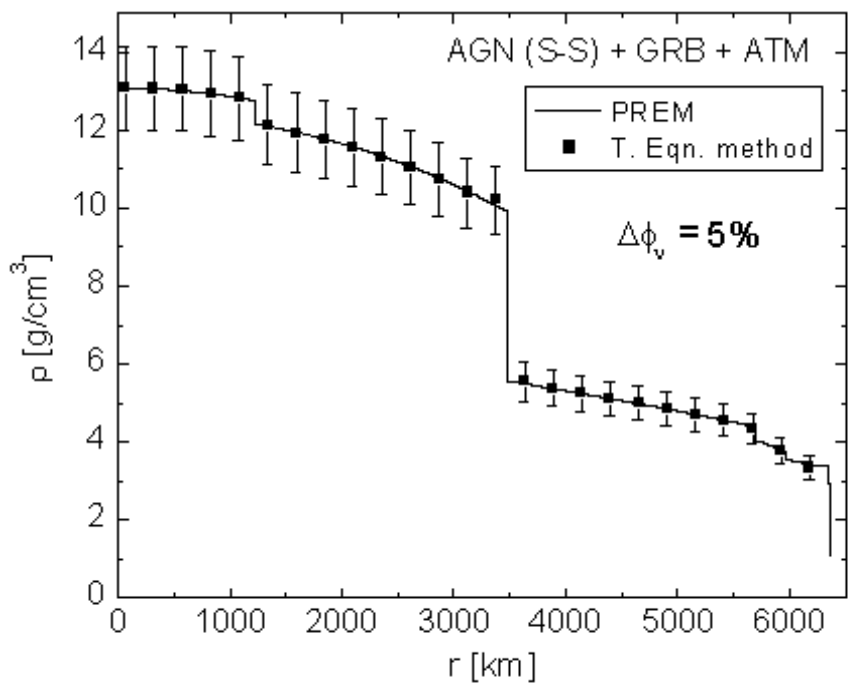
\noindent{\bf Figure 9:} Density recovered with error bars for an
uncertainty of $5\%$ in the surviving flux.

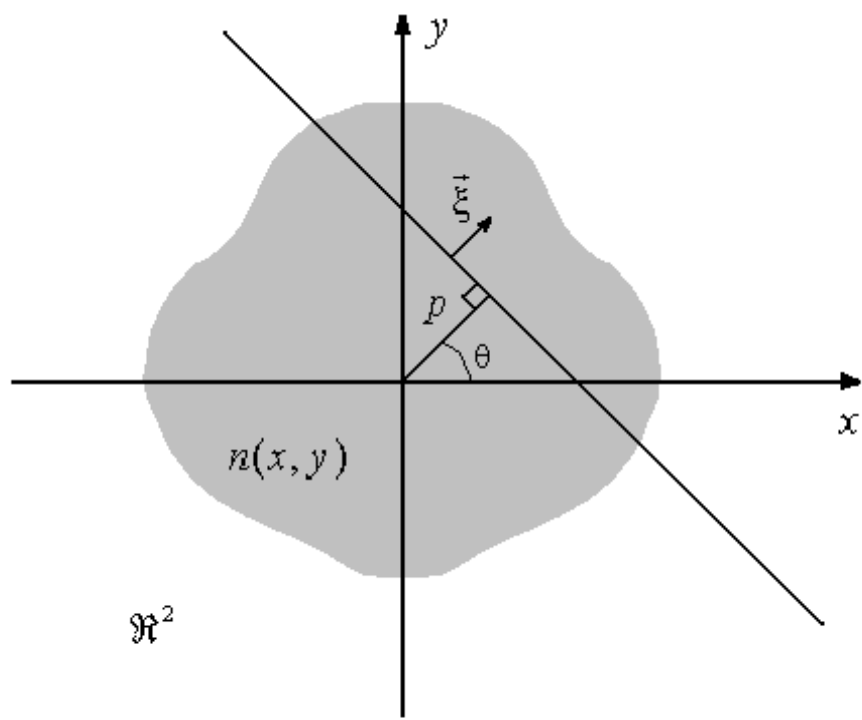
\noindent{\bf Figure 10:} ${\mathcal R}^2$ region where the Radon
transform is defined. The unit vector $\vec{\xi}$ and the
      number $p$ define the integration path.

\end{document}